\documentclass[usenatbib]{mnras}

\usepackage{natbib}
\usepackage{delarray}
\usepackage{color}
\usepackage{amsmath}
\usepackage{amssymb}
\usepackage{here}
\usepackage{graphicx}
\usepackage[utf8]{inputenc}
\usepackage{float}
\usepackage{epsfig}
\usepackage{tensor}
\usepackage{xspace}
\usepackage{multirow}

\usepackage{newtxtext,newtxmath}

\usepackage[T1]{fontenc}
\usepackage{ae,aecompl}

\hypersetup{draft}

\newcommand{\be}{\begin{equation}} \newcommand{\ee}{\end{equation}}

\newcommand{\solarmass}{\mathrm{M}_{\rm \sun}}
\newcommand{\solarluminosity}{\mathrm{L}_{\rm \sun}}

\newcommand{\msun}{M_{\odot}}

\newcommand{\acknowledgments}{\begin{small}\section*{Acknowledgments}\end{small}}
\newcommand\altaffilmark[1]{$^{#1}$}
\newcommand\altaffiltext[1]{$^{#1}$}

\title{Comparing Models for IMF Variation Across Cosmological Time in Milky Way-like Galaxies}

\author[Guszejnov, Hopkins, \&\ Ma]{
\parbox[t]{\textwidth}{ D\'avid Guszejnov\altaffilmark{1}\thanks{E-mail:guszejnov@caltech.edu}, Philip F. Hopkins\altaffilmark{1} and Xiangcheng Ma\altaffilmark{1}}
\vspace*{6pt} \\
\altaffiltext{1}{TAPIR, MC 350-17, California Institute of Technology, Pasadena, CA 91125, USA}
\vspace{-0.8cm}
}

\date{To be submitted to MNRAS, \today \vspace{-0.6cm}}
\begin{document}
\maketitle
\label{firstpage}

\begin{abstract}
One of the key observations regarding the stellar initial mass function (IMF) is its near-universality in the Milky Way (MW), which provides a powerful way to constrain different star formation models that predict the IMF. However, those models are almost universally ``cloud-scale'' or smaller -- they take as input or simulate single molecular clouds (GMCs), clumps, or cores, and predict the resulting IMF as a function of the cloud properties. Without a model for the progenitor properties of all clouds which formed the stars at different locations in the MW (including ancient stellar populations formed in high-redshift, likely gas-rich dwarf progenitor galaxies that looked little like the Galaxy today), the predictions cannot be fully explored, nor safely applied to ``live'' cosmological calculations of the IMF in different galaxies at different cosmological times. We therefore combine a suite of high-resolution cosmological simulations (from the Feedback In Realistic Environments project), which form MW-like galaxies with reasonable star formation properties and explicitly resolve massive GMCs, with various proposed cloud-scale IMF models. We apply the models independently to {\em every} star particle formed in the simulations to synthesize the predicted IMF in the present-day galaxy. We explore models where the IMF depends on Jeans mass, sonic or ``turbulent Bonner-Ebert'' mass, fragmentation with a polytropic equation-of-state, or where it is self-regulated by protostellar feedback. We show that all of these models, except the feedback-regulated ones, predict far more variation ($\sim 0.6-1$\,dex $1\,\sigma$ scatter in the IMF turnover mass) in the simulations than is observed in the MW.
\end{abstract}

\begin{keywords}
stars: formation -- turbulence -- galaxies: star formation -- cosmology: theory
\vspace{-1.0cm}
\end{keywords}

\section{Introduction}\label{sec:intro}

The (instantaneous) mass distribution of stars at their formation time, also known as the initial mass function (IMF), is one of the key predictions of any star formation model. This governs essentially all observable and theoretical aspects of star formation and stellar populations -- observable luminosities and colours; effects on stellar environments via feedback in the form of stellar winds, radiation, supernovae; nucleosynthesis and galactic chemical evolution, and so on. The IMF has been well-studied within the MW, and appears to be well-fit by a simple function with a \citet{Salpeter_slope} power-law slope at high masses and lognormal-like turnover at low masses \citep{Chabrier_IMF,Kroupa_IMF}. Perhaps the most interesting feature of the IMF is its universality: it has been found that there is quite weak variation within the MW \citep[for recent reviews, see][and references therein]{Chabrier_review_2003,IMF_review,IMF_universality,SF_big_problems}, albeit with a few possible outliers \citep[e.g.][]{Luhman_Taurus, Kraus_2017_Taurus}. As \citet{IMF_universality} emphasize, this universality includes both very young ($\sim$\,Myr-old) and very old ($\sim 10\,$Gyr-old) stellar populations; stars forming in small, nearby GMCs with masses $\sim 10^{4}-10^{6}\,\msun$ and massive complexes with masses $\sim 10^{6}-10^{7}\,\msun$; the solar neighbourhood at $\sim 10$\,kpc from the galactic centre (where the gas disk surface density is $\sim 10\,\msun\,{\rm pc^{-2}}$) and the central molecular zone at sub-kpc and $\sim100\,$pc scales (where gas surface densities are order-of-magnitude larger). 

In other galaxies, the IMF usually must be {\em assumed}, and with an IMF assumption, physical properties of the stellar populations and galaxies (e.g.\ their stellar masses) are derived from observables (e.g.\ light, colours). This makes it critical to understand the IMF, in order to understand galaxy formation. Likewise it is critical for models of galaxy formation to predict or assume {\em some} IMF model, in order to make any meaningful predictions for observable quantities. The universality of the IMF in old stellar populations in the MW is widely taken as a suggestion that it may be near-universal in other galaxies, because older populations in the MW formed when the galaxy was much younger and very different, likely a typical high-redshift gas-rich, metal-poor dwarf galaxy. There are indirect constraints on the IMF both from spectral features and integrated mass-to-light constraints in nearby galaxies: these mostly also favor a universal IMF \citep[e.g.][]{Fumagalli_2011_universal_IMF,Koda_2012_univseral_IMF,Andrews_2013_survey,Andrews_2014_survey,Weisz_2015_survey}. More recently there have been more interesting hints of variation in the centres of massive elliptical galaxies \citep{Conroy_vanDokkum_ellipticals,vanDokkum_conroy_2011,ConryVanDokkum_IMFvar_2012,Treu_galaxy_IMF_2010,Sonnenfeld_Treu_2015_IMF,Cappellari_IMF_var_2012,Posacki_Cappellari_2015_IMF,Navarro_2015_IMF_z,Navarro_2015_IMF_relic,Navarro_2015_IMF_relation}, and perhaps also in faint dwarf galaxies \citep{Hoversten_Glazebrook_dwarf_IMF_2008,Brown_dwarf_galaxy_2012,Geha_turnover_univ}. Even so, it is worth stressing that the implied variation is not radical: it implies variation of a factor $<2$ in the stellar mass-to-light ratio. 

As a result, there is a long history of both theoretical and empirical models for galaxy formation which have attempted to predict the IMFs that should arise in different galaxy populations, as a function of either galaxy-scale or $\sim$kpc-scale properties (what can be resolved in most previous calculations) within the galaxies \citep[see e.g.][]{Baugh2005,vanDokkum2008_top_heavy_IMF,Hopkins_CMF_var,Dave_2012_galaxy_evol,Narayanan_2012_jeans_extragalactic,Recchi2015IGIMFmodel,Lacey2016galaxyIMFvariationModel,Blancato_2016_Illustrus_IMF_var}. At the same time, the specific hints of galaxy-to-galaxy variation discussed above have prompted a new wave of theoretical models which argue the IMF could vary under certain conditions at the GMC or sub-GMC scale, in a way that may connect to the systematic variation inferred in different galaxies (e.g. \citealt{Weidner_2013,Bekki_2013,Chabrier_hennebelle_2014,Ferreras_Weidner_2015}). 

However, these models in every case rely on very strong simplifying assumptions -- the IMF is predicted as a function of the cloud properties out of which the stars form, such as its temperature, density, turbulent velocity dispersion, virial parameter, etc. (from which properties like the Jeans mass, or the turbulent Bonner-Ebert mass, or the IGIMF turnover mass, are determined). It is impossible at present to know these empirically because all the clouds that formed the old stellar populations in a galaxy (or even for most clouds within a galaxy at present day), so instead some strong additional assumptions are usually applied. For example, authors assume isothermal gas with $T=10\,$K (or some other temperature) at all densities, a universal linewidth-size relation across all galaxies, redshifts, and regions within galaxies, or a Jeans or sonic mass within clouds that somehow scales proportionally to that measured from the gas at the $\gtrsim $kpc scales resolved in the cosmological calculations. But if these properties vary across cosmic time, or cloud-to-cloud, then any such model will produce variation in the predicted IMF which can be compared to the observational limits within the MW. 

In this paper we therefore investigate the predicted variation in the IMF peak imprinted by these physics in a number of IMF models. We combine high-resolution simulations of MW-like galaxies (where the cloud-scale properties can be at least partially resolved) with the relevant small-scale models for IMF variation as a function of cloud properties. 

\begin{figure}
\begin {center}
\includegraphics[width=\linewidth]{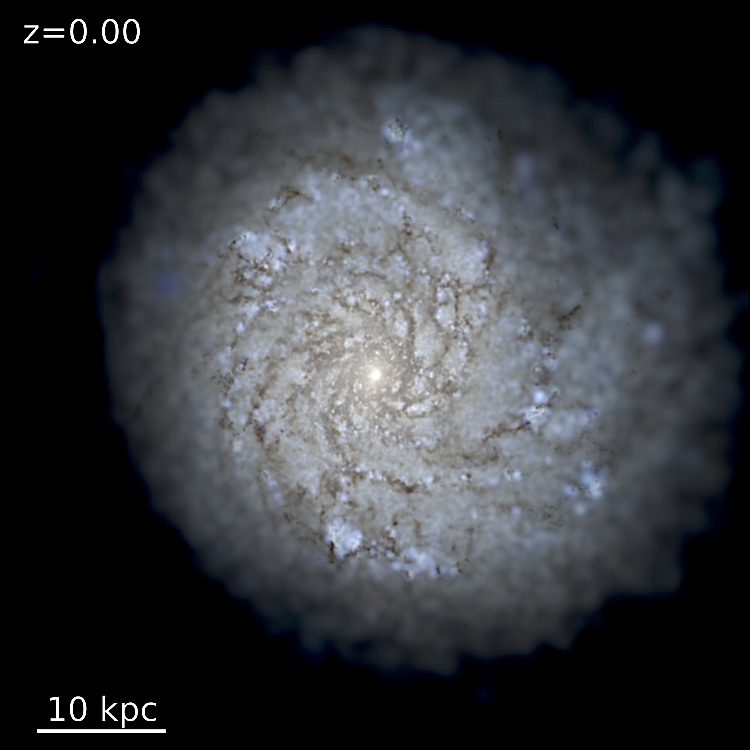}
\vspace{-0.5cm}
\caption{Visualization of the starlight (mock \textit{ugr} composite image, accounting for each stellar sink-particle's age and metallicity and ray-tracing including dust obscuration) from one of the simulated MW-like galaxies (see Table~\ref{tab:SF_models}) we use in our calculations (galaxy {\bf m12i} from \citealt{Hopkins2017_FIRE2} with $56000,\solarmass$ resolution). Note that resolved molecular clouds and arms are evident. See Fig. \ref{fig:all_galaxy_views} for the other galaxies from Table \ref{tab:SF_models}.}
\label{fig:galaxy_fig}
\vspace{-0.5cm}
\end {center}
\end{figure}

\begin{figure*}
\begin {center}
\includegraphics[width=0.33 \linewidth]{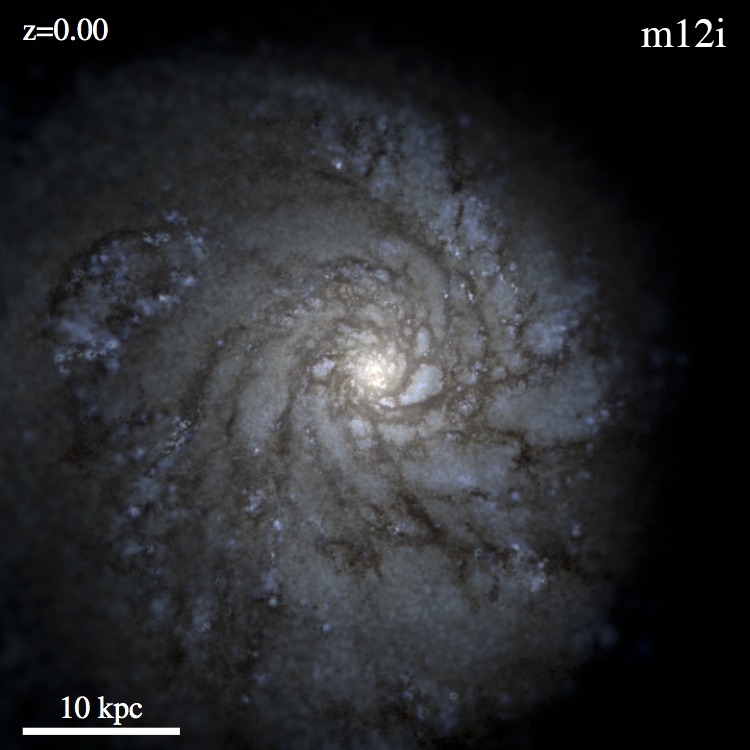}
\includegraphics[width=0.33 \linewidth]{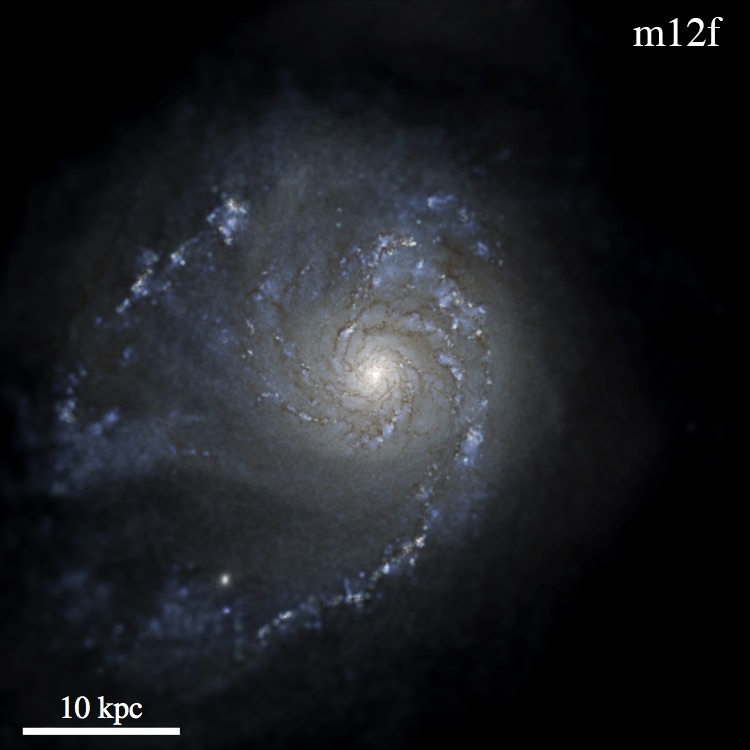}
\includegraphics[width=0.33 \linewidth]{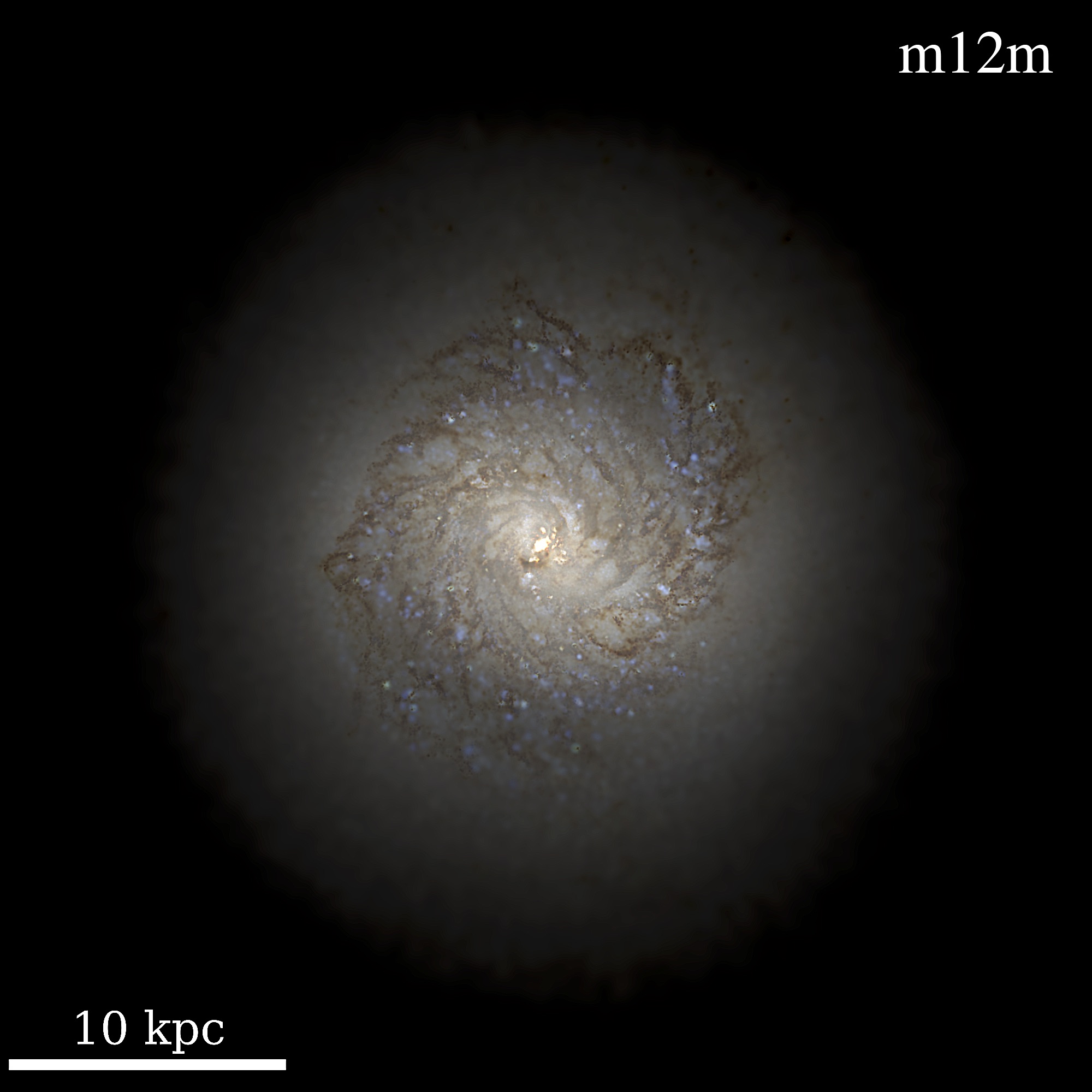}
\\
\includegraphics[width=0.33 \linewidth]{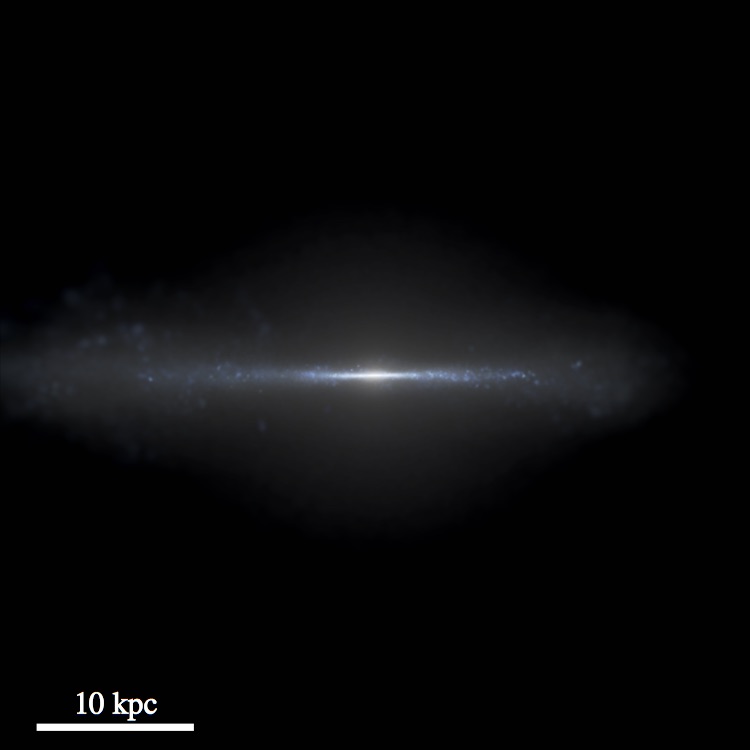}
\includegraphics[width=0.33 \linewidth]{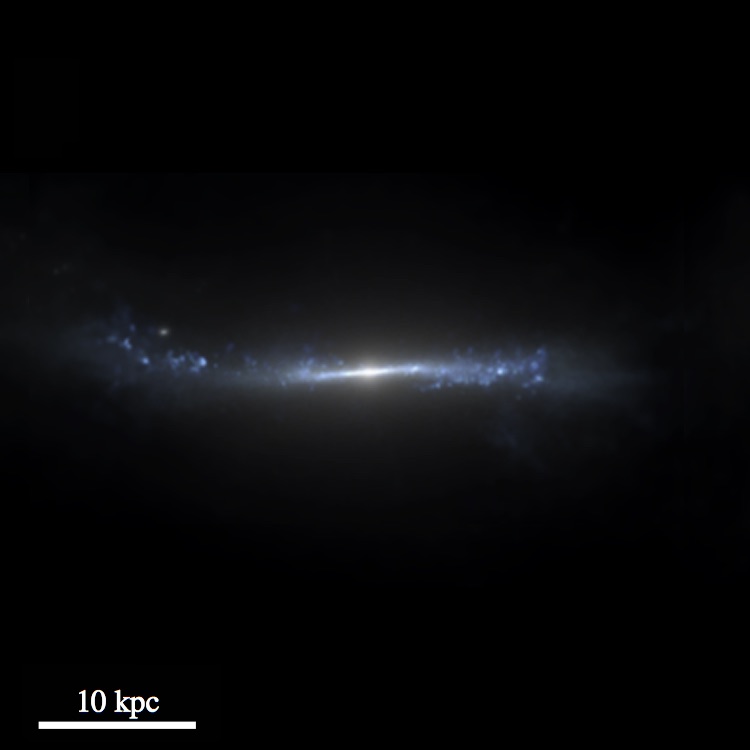}
\includegraphics[width=0.33 \linewidth]{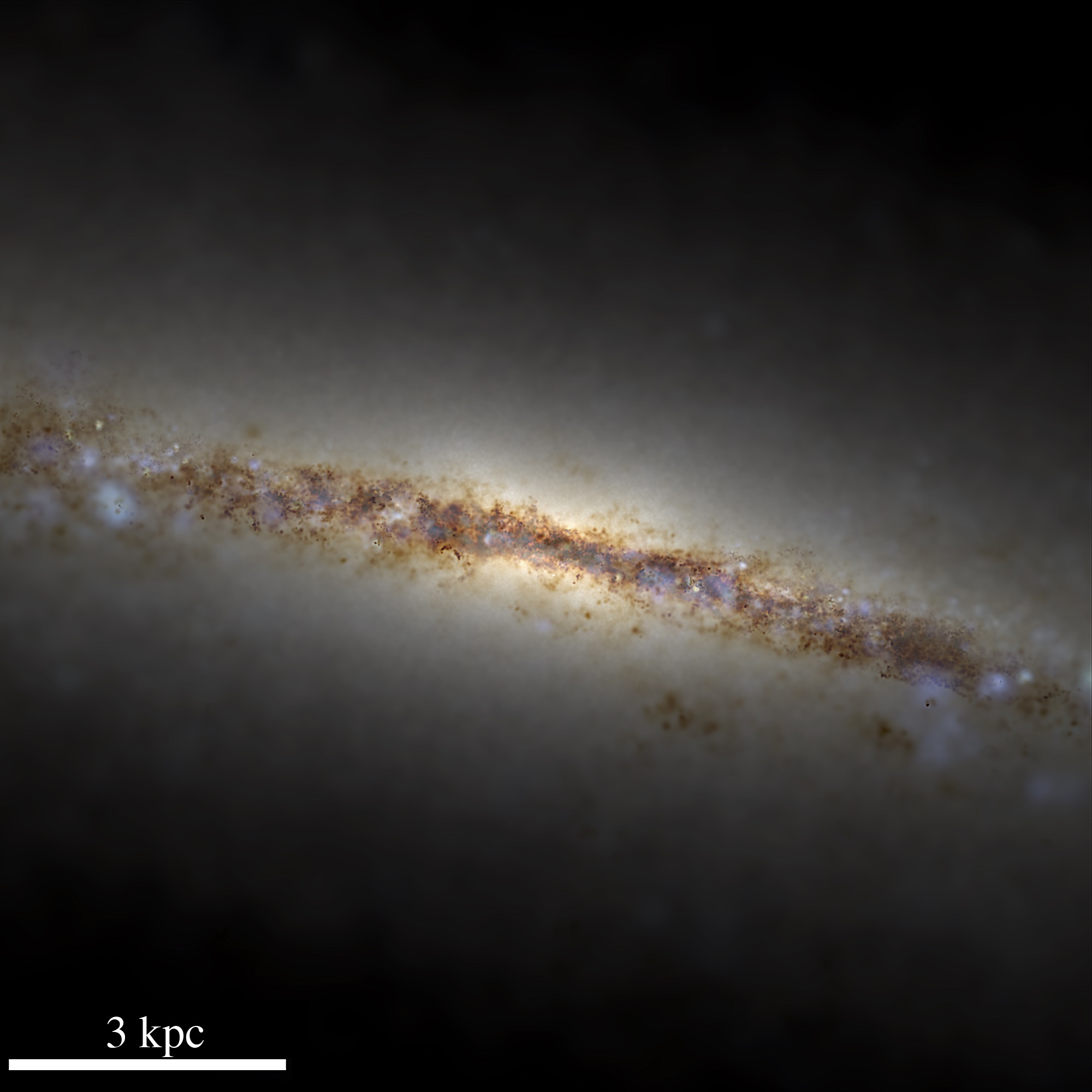}

\caption{Face on (top) and edge on (bottom) visualizations of starlight (mock \textit{ugr} composite image, accounting for each stellar sink-particle's age and metallicity and ray-tracing including dust obscuration) for the simulated MW-like galaxies from Table~\ref{tab:SF_models}. Unlike Fig. \protect\ref{fig:galaxy_fig} the m12i example shown here is from the high resolution ($7000\,\solarmass$) run. The MHD run is not shown as it gives virtually identical results as the non-MHD runs \citep[see][]{KungYi_weak_MHD_2016}. Note the edge-on images of {\bf m12i} and {\bf m12f} are a mock Galactic (Aitoff) projection from a random star at $\sim10\,$kpc from the galactic center. For more details on the individual runs see \citealt{Hopkins2017_FIRE2}.}
\label{fig:all_galaxy_views}
\end {center}
\end{figure*}

\begin{figure}
\begin {center}
\hspace{-0.1cm}\includegraphics[width=1.05 \linewidth]{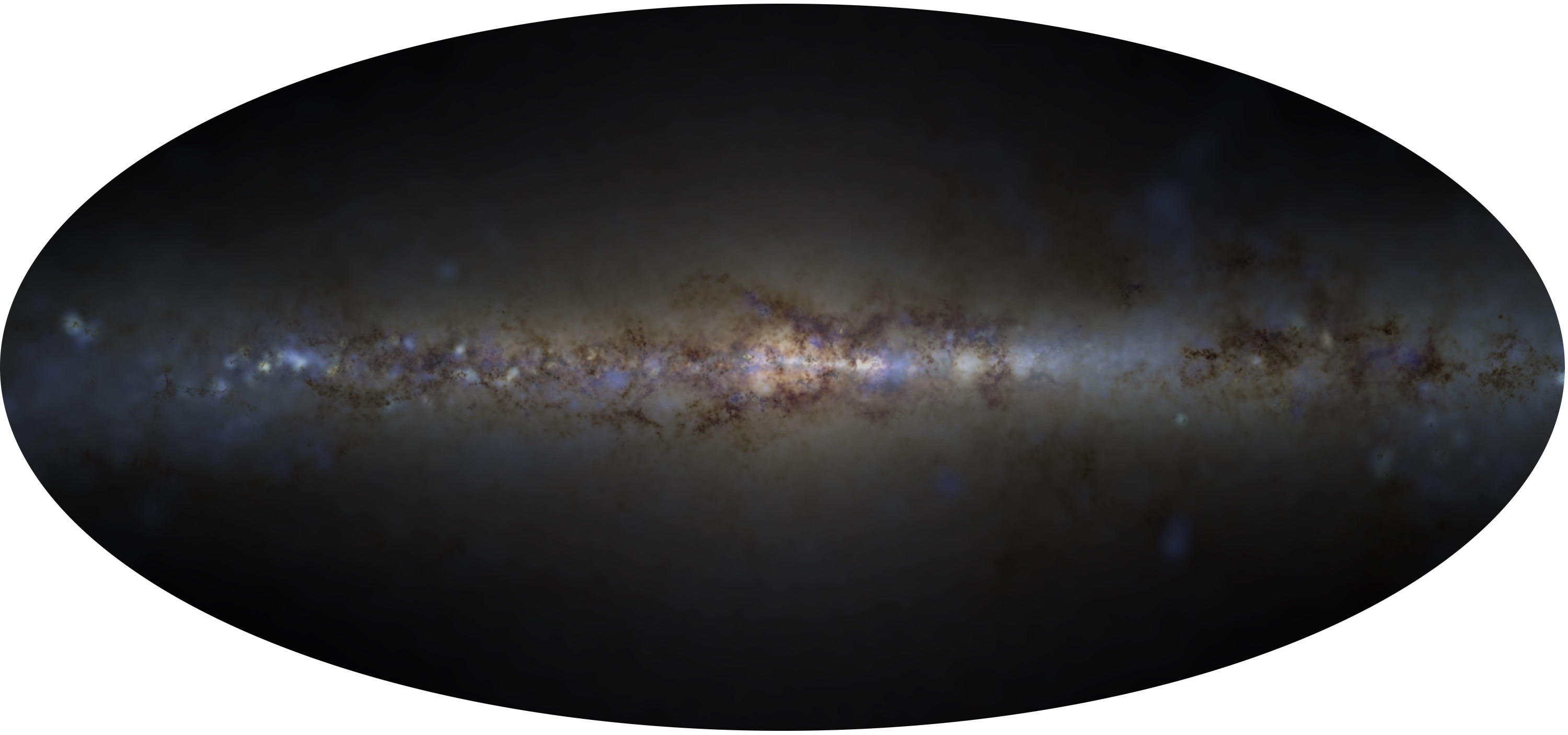}\\
\hspace{-0.1cm}\includegraphics[width=1.05 \linewidth]{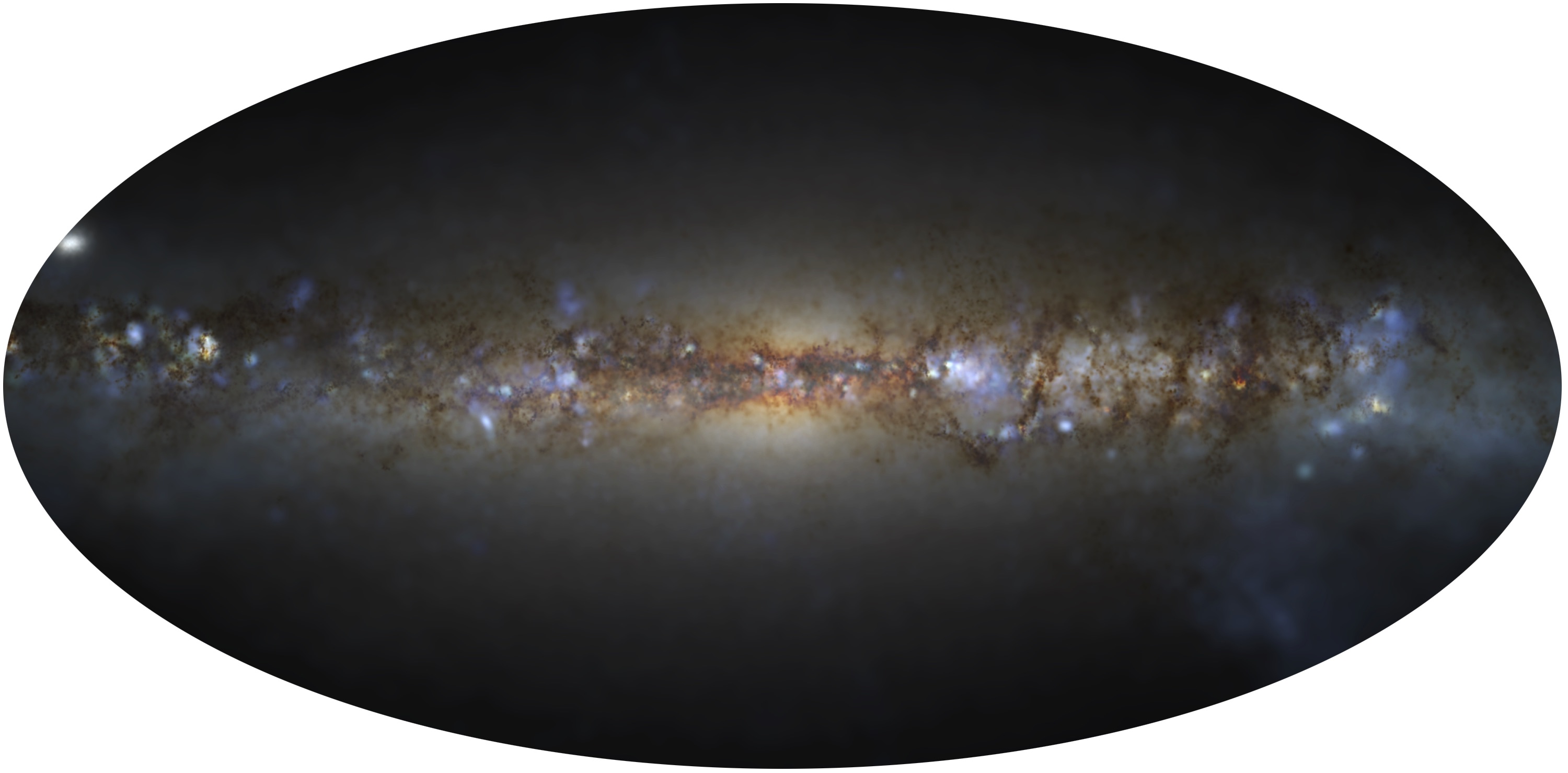}\\
\caption{Mock Galactic (Aitoff) projection from a random star at $\sim10\,$kpc from the galactic center, for {\bf m12i} ({\em top}) and {\bf m12f} ({\em bottom}). For more details on the individual runs see \citealt{Hopkins2017_FIRE2}.}
\label{fig:all_galaxy_views_aitoff}
\end {center}
\end{figure}

\vspace{-0.5cm}
\section{Model and Methodology}\label{sec:method}

\subsection{Simulation}

We utilize a set of numerical simulations of MW-like galaxies (see Table~\ref{tab:SF_models}) presented in \citet{Hopkins2017_FIRE2}, from the  Feedback  in  Realistic  Environments  (FIRE) project (\citealt{Hopkins2014_FIRE}).\footnote{\url{http://fire.northwestern.edu}} The simulations are fully cosmological ``zoom-in'' simulations (where resolution is concentrated on one galaxy in a large cosmological box, run from redshift $z>100$ to today) and are run using GIZMO (\citealt{Hopkins2015_GIZMO})\footnote{\url{http://www.tapir.caltech.edu/~phopkins/Site/GIZMO.html}}, with the mesh-free Godunov ``MFM'' method for the hydrodynamics \citep{Hopkins2015_GIZMO}. Self-gravity is included with fully-adaptive force and hydrodynamic resolution; the simulation mass resolution is fixed at $7000$ or $56000\,\msun$ (Table~\ref{tab:SF_models}). 
The simulations include detailed metallicity-dependent cooling physics from $T=10-10^{10}\,$K, including photo-ionization/recombination, thermal bremsstrahlung, Compton, photoelectric, metal line (following \citealt{Wiersma2009_cooling}), molecular, fine structure (following \citealt{CLOUDY}), dust collisional and cosmic ray processes, including both a meta-galactic UV background and each star in the simulation as a local source. 

Individual stars are not resolved in the simulations; but star formation is approximated from resolved scales via a sink-particle method. Gas which is locally self-gravitating, self-shielding, Jeans unstable, and exceeds a minimum density $n>n_{\rm crit}=1000\,{\rm cm^{-3}}$ (Table~\ref{tab:SF_models}) is transformed into ``star cluster sink particles'' on its dynamical time. Each such particle represents an IMF-averaged single stellar population of the same age and metallicity, with mass equal to the mass resolution. 

Once formed, the simulations include feedback from these star particles via OB \&\ AGB mass-loss, SNe Ia \&\ II, and multi-wavelength photo-heating and radiation pressure; with inputs taken directly from stellar evolution models \citep{1Leitherer_1999_Starburst99}, assuming (in-code) a universal IMF \citep{Kroupa_IMF}.

There are two reasons for using cosmological simulations instead of present-day observational data or a more localized cloud simulation. (1) We wish to test and validate the approach of using cloud-scale IMF models {\em dynamically} in next-generation cosmological simulations. Because stellar feedback and observable properties depend on the IMF, truly self-consistent predictions should include some sub-grid IMF model. These cosmological simulations were run assuming a universal IMF, but others (see references in \S~\ref{sec:intro}) adopt a dynamical IMF model based on resolution-scale properties. But it has not been asked whether the models they considered violate observational constraints in the MW. (2) Stars at a given present-day position in a galaxy can form at wildly different times/places (some even in other dwarf galaxies). This is especially true for the stars in old MW clusters, which appear to have formed at high redshifts, probably in distinct dwarf galaxies with entirely distinct radiation fields, turbulent velocity dispersions, gas masses, etc. It is impossible to know the distribution of progenitor cloud properties {\em at star formation for old stars} (needed for a given IMF model to make predictions) from observations (or localized simulations) alone. 

The galaxies studied here, shown in Figs.~\ref{fig:galaxy_fig}-\ref{fig:all_galaxy_views} -- have been studied extensively in previous work: they are similar to the MW in their stellar mass, present-day gas fractions and SFRs \citep{Hopkins2014_FIRE}, and metallicity \citep{Ma_FIRE_metallicity}. Our ``fiducial case'' {\bf m12i} is also similar to the MW in its stellar kinematics, thin+thick disk morphology, metallicity gradient and metal abundance ratio gradients (in both vertical and radial directions) stellar age distribution \citep{Ma_thin_disk,Ma_2017_metallicity}, $R$-process element distribution \citep{van_de_Voort_r_process} and galactic stellar halo and satellite dwarf population \citep{Wetzel_LATTE_2016}. The other two examples represent a slightly later-type ({\bf m12f}) and earlier-type ({\bf m12m}) galaxy, at the same stellar mass and SFR. This is particularly useful because of course no simulation will exactly match the formation history of the MW, so it is important to understand whether our predictions are sensitive to this.

\begin{figure}
\begin {center}
\includegraphics[width=\linewidth]{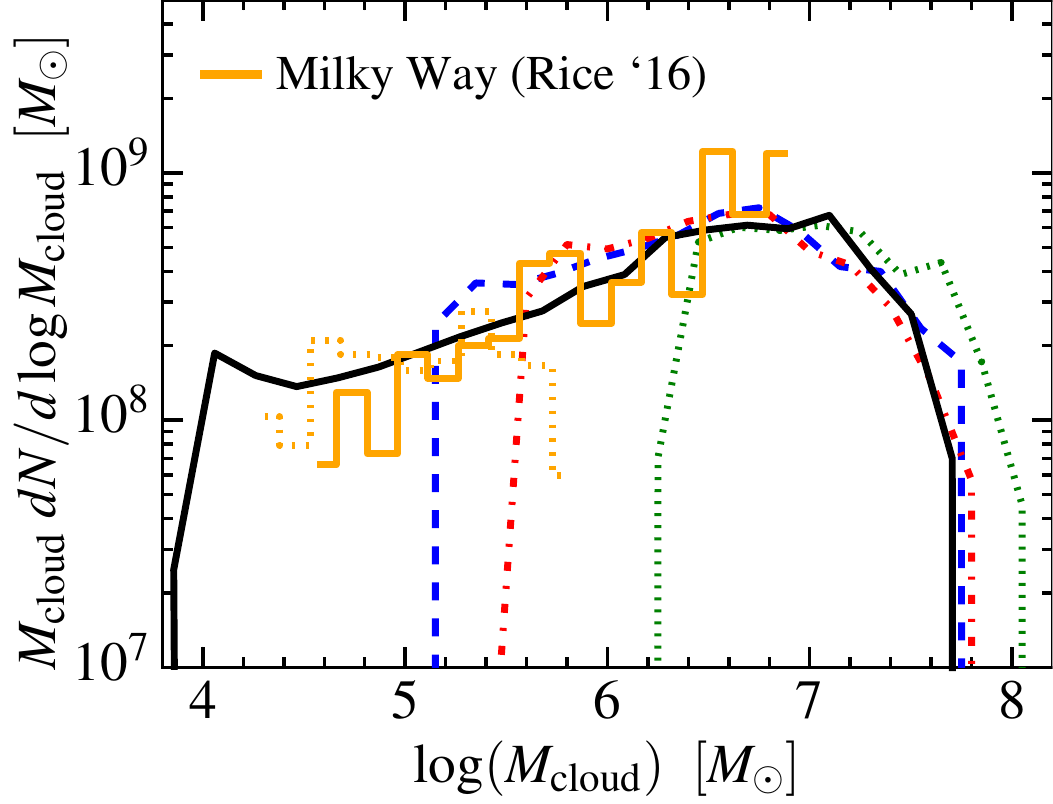}\\
\includegraphics[width=\linewidth]{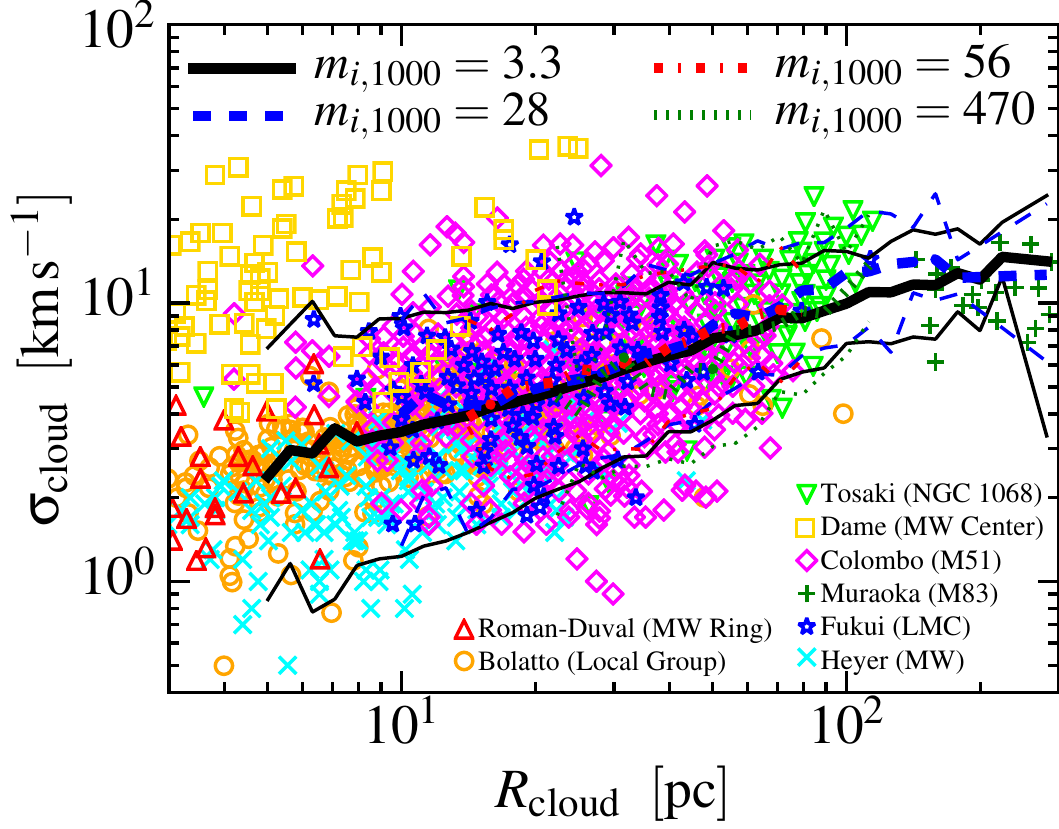}
\vspace{-0.5cm}
\caption{{\em Top:} Mass function (MF) of GMCs in the fiducial simulation (Fig.~\ref{fig:galaxy_fig}) at $z\approx0$, at different mass resolution (baryonic particle mass $m_{i}=1000\,m_{i,\,1000}\,\msun$). We restart the fiducial simulation from Fig.~\ref{fig:galaxy_fig} (with $m_{i,\,1000}=56$) at $z=0.1$, after re-sampling the particles to raise/lower the mass resolution. We then evolve it for $\sim 1\,$Gyr to $z=0$, and measure the MF of dense cold-gas clouds (identified in post-processing with a friends-of-friends group-finder) time-averaged over the last $\sim 100\,$Myr inside $<20$\,kpc of the galaxy center. All details of the resampling and group-finding method are in \citet{Hopkins2017_FIRE2}. We compare the observed MW GMC MF from \citet{Rice2016_MW_GMC_Catalogue}, normalized to the same total mass, measured inside ({\em solid}) and outside ({\em dotted}) the solar circle. At all resolutions, a GMC MF similar to that observed is recovered. The most massive GMCs contain most of the mass/star formation and are the first-resolved. At higher resolution we extend to smaller GMCs.
{\em Bottom:} Linewidth-size relation for the same clouds (median in thick lines; $5-95\%$ intervals in thin lines), vs.\ observations in nearby galaxies (\citep{Bolatto_2008,Fukui_2008_LMC_survey,Heyer_2009_Larson,Muraoka_2009_M83,RomanDuval_2010_clouds,Colombo_2014_PAWS_survey,Heyer_Dame_2015,Tosaki_2017_linewidth_size_data}; note our definition of $R_{\rm cloud}$ is equivalent to their $\sigma_{r}$). The predicted normalization and $1\,\sigma$ dispersion ($\approx 0.12$\,dex, although it increases slightly to $\approx 0.2$ dex at the lowest masses) are consistent with observations (compare e.g.\ \citealt{Kauffmann_Pillai_2013}). There is no systematic resolution dependence (other than sampling smaller clouds at higher resolution).\vspace{-0.5cm}}
\label{fig:mass_functions}
\vspace{-0.5cm}
\end {center}
\end{figure}

These and other FIRE simulations have also been shown to reproduce the observed Kennicutt-Schmidt relation \citep{Sparre_FIRE_SFR_2015,Orr_FIRE_KS}, properties of galactic outflows \citep{Muratov_FIRE_winds_2015} and (in higher resolution, non-cosmological simulations) the observed mass function (and CO luminosities), size-mass, and linewidth-size distributions of GMCs \citep{Hopkins_2012_galaxy_structure,Hopkins_2013_dense_gas}. One might reasonably worry that this cannot be captured at the lower resolution necessary in cosmological simulations. Therefore in Fig.~\ref{fig:mass_functions} we plot the mass function and linewidth-size relation of GMCs identified at present-day in the actual simulations studied here. They appear to agree at least plausibly with observed properties \citep{Dobbs_2014,Heyer_Dame_2015}. Note that our mass resolution introduces a cutoff at the low mass end of the GMC mass function because these clouds can not be resolved by the simulation. However, all simulations included in this paper do resolve the most massive GMCs ($>10^6\,\solarmass$), in which most of the mass is concentrated (owing to the shape of the GMC mass function), allowing us to recover galactic properties even at lower resolutions. This is clearly apparent in the linewidth-size relation which shows a good agreement with \cite{Bolatto_2008}. All of this is not to say that the simulations are perfect analogues to the MW; however they are at least a reasonable starting point (see Fig. \ref{fig:mass_functions} for details). 

Our choice of $n_{\rm crit}=1000\,{\rm cm^{-3}}$ (at mass resolution $7000-56000\,\msun$) can be justified by assuming that GMCs follow the mass-size relation of \cite{Bolatto_2008} ($M_{\rm cloud}\sim \pi\,(85\,\msun\,{\rm pc^{-2}})\,R_{\rm cloud}^{2}$), which implies the density threshold for star formation is slightly higher than the mean density of the most-dense resolved clouds. More specifically, we chose the density threshold to correspond to the typical density where the Jeans/Toomre fragmentation scale falls below our mass resolution. In either case the GMC mass function and SFR is dominated by the most massive (hence well-resolved) clouds. This is evident in Fig.~\ref{fig:mass_functions}, where we show the GMC mass function and linewidth-size relation predicted at present-day in the galaxy both (a) agree reasonably well with observations (within a factor $\sim 2$ at all cloud sizes/masses resolved), and (b) are insensitive to resolution (except, of course, that at higher resolution they extend to smaller GMCs). This gives us some confidence that our predictions are not strongly resolution-dependent. In Table~\ref{tab:SF_models} we show that varying resolution and physics (in an otherwise identical run including magnetic fields, {\bf m12i+MHD}) do not significantly alter our predictions.

\begin{figure}
\begin {center}
\hspace{-0.2cm}\includegraphics[width=\linewidth]{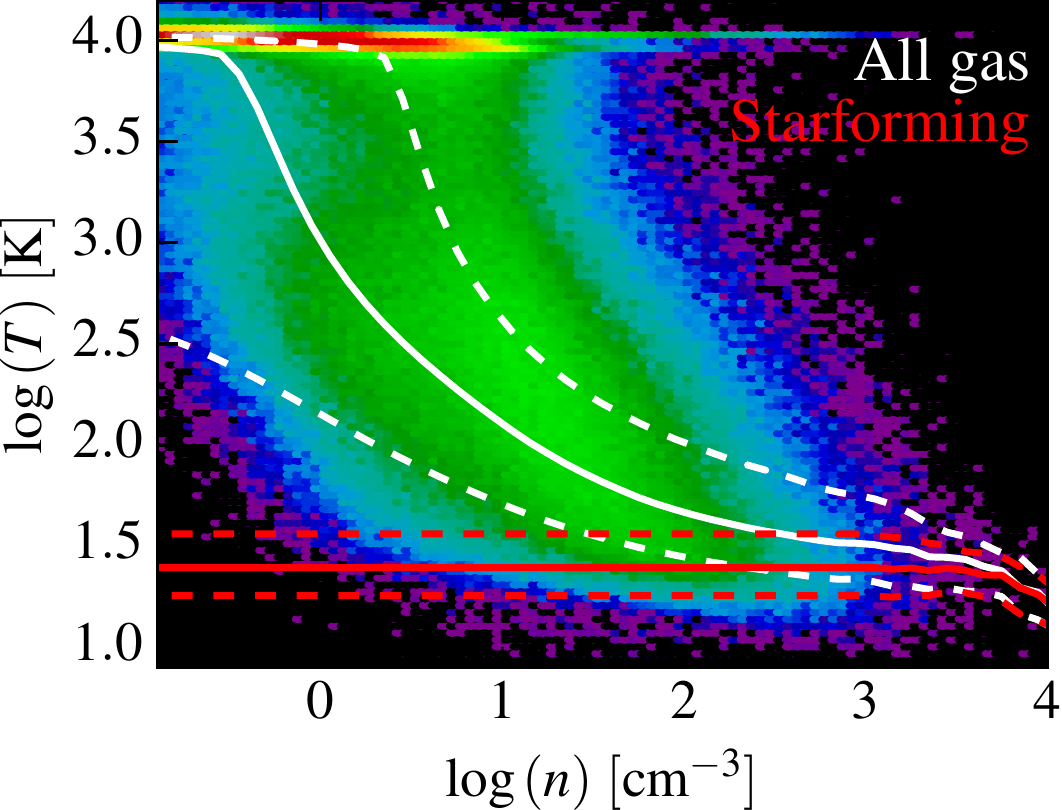}\\
\vspace{-0.3cm}
\caption{
Density-temperature diagram for gas at {\em present-day} ($z=0$) in our high-resolution {\bf m12i} simulation (others are similar). Colors show a 2D histogram colored by the gas-mass per-pixel (log-weighted, increasing black-blue-green-red with a $\sim 6$\,dex stretch), so this is peaked where there is significant mass in a narrow temperature range. HII regions, warm ionized medium, and warm and cool neutral phases are evident (we do not show lower densities where hot gas is prevalent). Solid (dashed) lines show the median (inter-quartile) temperature of all gas denser than $>n$, weighted by mass ({\em white}) or star formation rate ({\em red}). The latter converges rapidly because SF is restricted to high-$n$ gas. At $z=0$ in the simulation, most SF occurs in gas with $n> 1000\,{\rm cm^{-3}}$ and $T\approx 20-30\,$K.
}
\label{fig:n_T_hist_z0}
\vspace{-0.5cm}
\end {center}
\end{figure}

\begin{figure}
\begin {center}
\hspace{-0.2cm}\includegraphics[width=\linewidth]{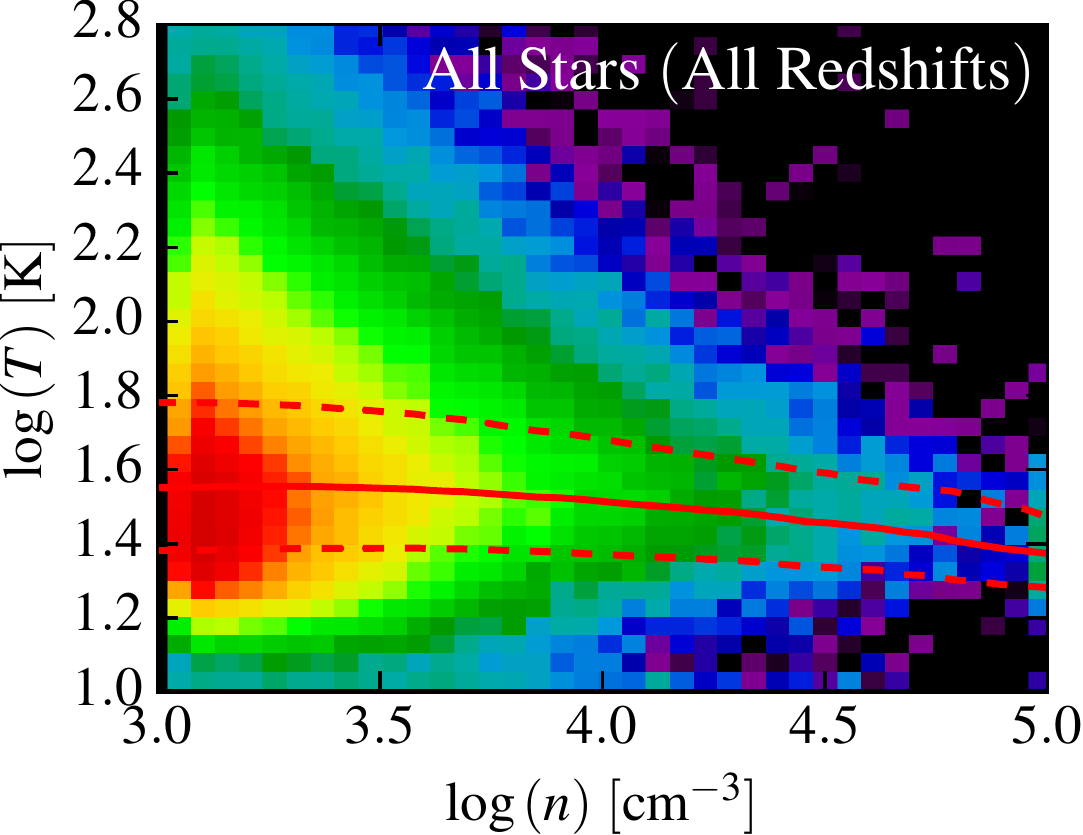}
\vspace{-0.3cm}
\caption{Density-temperature diagram (as Fig.~\ref{fig:n_T_hist_z0}; same galaxy), but for the progenitor clouds (gas elements which formed a stellar sink particle) of {\em all} stars which reside in the $z=0$ galaxy (integrated over all cosmic time). Note these are the cloud properties {\em at the moment the sink formed}, weighted by mass in stars today (colors use $\sim3\,$dex stretch). As expected, most sinks form a factor of a few above our minimum threshold ($n_{\rm crit}=1000\,{\rm cm^{-3}}$), though some gas reaches much higher densities. Lines again show the median and inter-quartile range for stars formed at resolved densities $>n$. Accounting for different times and progenitor galaxies, the dispersion in temperatures at a given density is a factor $\sim 3-4$ larger here than for star-forming gas just at $z=0$.}
\label{fig:n_T_hist}
\vspace{-0.5cm}
\end {center}
\end{figure}

Because the simulations resolve down to cloud scales, but no further, we treat each star-forming gas element as an independent ``parent cloud'', which sets the initial conditions for its own detailed IMF model (in accordance with the IMF models we investigate). Specifically, whenever a sink particle is spawned, we record all properties of the parent gas element from which it formed, and use these in post-processing to predict the IMF. Fig.~\ref{fig:n_T_hist_z0} shows the properties of gas elements at one instant, $z=0$, weighted by star formation rate. Integrating over all times and all galaxies which form stars that ultimately reside in the final galaxy, Fig~\ref{fig:n_T_hist} shows the density and temperature distribution of these ``star forming particles'' (gas at the moment the simulation assigned its mass to a sink particle) at the time of their formation, from our high-resolution {\bf m12i} run. Not surprisingly most sinks form around the simulation density threshold from this particular run ($\sim 1000\,{\rm cm^{-3}}$). This choice has no effect on the scatter in both $n_{\rm cloud}$ and $T_{\rm cloud}$, which are the relevant parts to our study. Scatter in these quantities translate to a scatter in the local IMF according to the IMF models we are studying. Note that \citet{Narayanan_2013_XCO} show the inferred temperature range from mock CO-ladder observations will tend to be significantly smaller than the range plotted here. We wish to emphasize that what is plotted in Fig. \ref{fig:n_T_hist} is not the density/temperature of the core or proto-stellar gas that which directly collapses and forms stars; that is not resolved in these simulations. Instead these are the properties of the progenitor molecular clouds, measured at the smallest resolved scales, which will (and must, physically) fragment into denser sub-clumps that can directly form stars. Also, the width in this distribution is expected to be higher than in present day clouds because of the longer lifetime of stars which preserves the progenitor cloud properties in their IMF for cosmological timescales.

\vspace{-0.5cm}
\subsection{From Parent Cloud to IMF Properties}
\label{sec:methods.imf.models}

\begin{figure}
\begin {center}
\includegraphics[width=\linewidth]{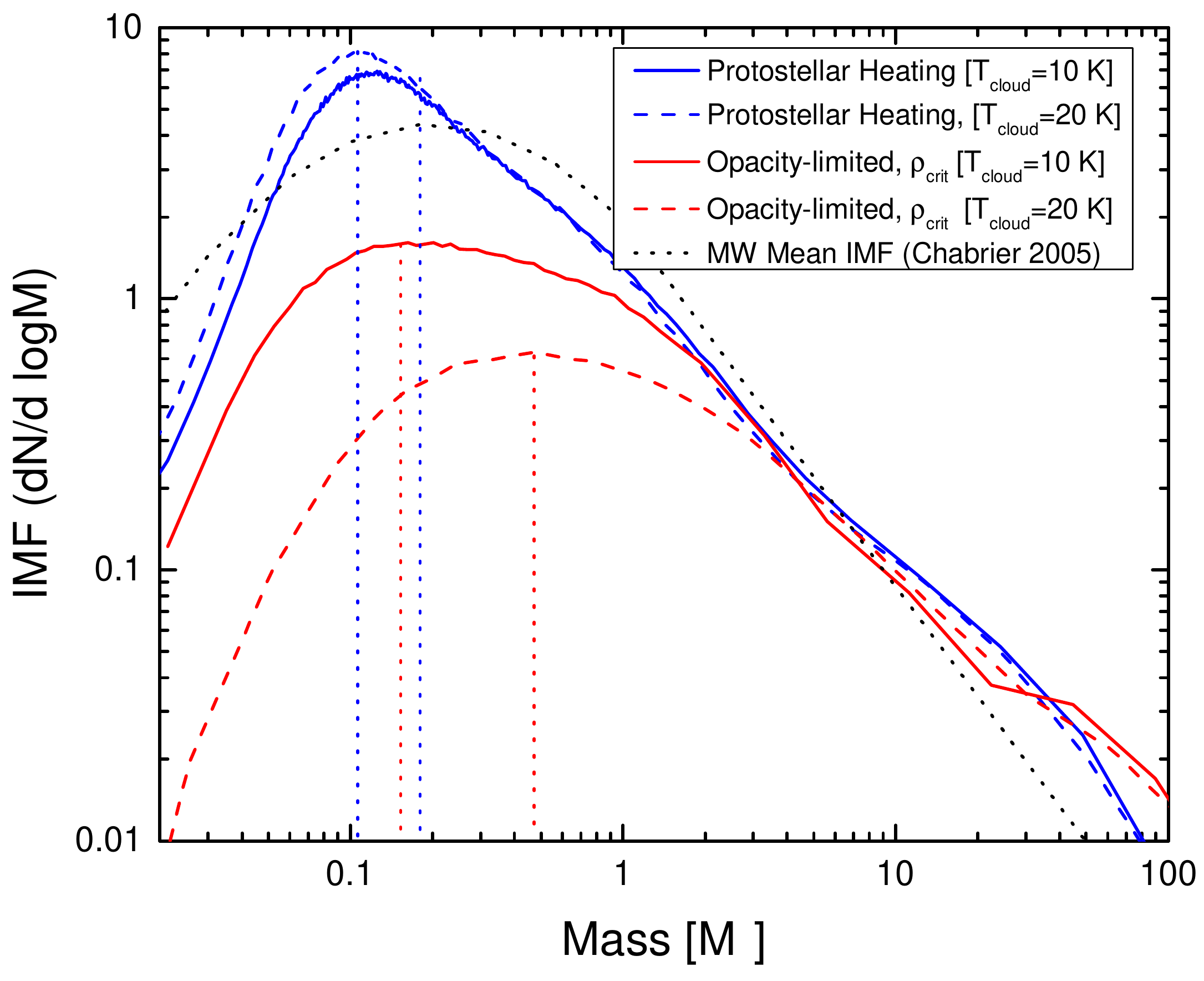}
\vspace{-0.5cm}
\caption{Predicted IMF using the framework of \citet{guszejnov_GMC_IMF}, within progenitor clouds with different initial temperatures $T_{\rm cloud}=10\,$K or $20\,$K. We compare two IMF models from Table~\ref{tab:SF_models}: (1) accounting for proto-stellar heating, and (2) ignoring heating and treating the gas with a polytropic equation-of-state until some it reaches the opacity limit. We compare the standard fit to the observed IMF from \citet{Chabrier_IMF}. Differences in temperature produce different model shifts, per the scalings in Table~\ref{tab:SF_models}.}
\label{fig:IMF_example}
\vspace{-1cm}
\end {center}
\end{figure}

From this point we infer the IMF turnover mass from the initial conditions of these parent clouds. This exercise has been done in detail by \cite{guszejnov_feedback_necessity} where the semi analytical framework of \cite{guszejnov_GMC_IMF} was utilized to create a mapping between GMC properties and the IMF. Fig. \ref{fig:IMF_example} shows how the IMF peak scales with initial temperature in an equation of state (EOS) and a protostellar feedback based IMF model. Such scaling relations can be analytically derived for other IMF models (e.g. Jeans mass) as well -- we focus here on how each model predicts the turnover or ``critical'' mass $M_{\rm crit}$ scale, because this is the most identifiable feature of the IMF (it sets the mass-to-light ratio, and varies significantly between models). In contrast the bright-end slope varies negligibly between models\footnote{Note that observations do indicate variations in the IMF slopes in extra-galactic populations \protect\citep[e.g.][]{Cappellari_IMF_var_2012, Spiniello_2012_IMF_steepen, Shu_2015_survey} but these measurements only sample the relatively low mass region of the IMF ($<\solarmass$). The IMF in the MW, however, is well sampled at higher masses and appears to be consistent with a near-universal power-law tail \citep{IMF_universality}, which most IMF models are able to roughly reproduce (see references in Table \ref{tab:SF_models} for specifics in each case)}, so it is not useful as a diagnostic.

In this paper we investigate the sensitivity to initial conditions for the following classes of IMF models (summarized in Table~\ref{tab:SF_models}):
\begin{itemize}
  \item \textbf{Jeans mass models:} The Jeans instability is the primary mechanism for the collapse of gas clouds into stars, so these models assume that IMF properties are set by local mean Jeans mass of the parent molecular cloud complex (e.g. \citealt{BateBonell2005}). Therefore, the critical mass is
	\be
	M_{\rm crit,J}\sim \frac{\pi c_s^3}{6 G^{3/2}\rho^{1/2}}.
	\ee
    Note that the models may still assume sub-fragmentation to smaller scales, but the key assumption (for our purposes) is simply that the turnover mass somehow scales proportional to the parent cloud Jeans mass.
	
	\item \textbf{Opacity limit equation of state (EOS) models:} As the molecular gas becomes denser it reaches the point where it becomes opaque to its own cooling radiation, leading to a transition from isothermal to adiabatic behavior, terminating fragmentation at the Jeans mass at this density. This can occur at a critical volume density $\rho_{\rm crit}$ (e.g. \citealt{LowLyndenBell1976,Whitworth98a,Larson2005,Glover_EQS_lowgamma_ref, Jappsen_EQS_ref, Masunaga_EQS_highgamma_ref}). Motivated by radiation transfer simulations like \citealt{Bate_2009_rad_importance} we also investigated the case where the transition occurs at a critical surface density $\Sigma_{\rm crit}$. The critical masses in these cases are:
	\begin{eqnarray}
	M_{\rm crit,\rho}\sim \frac{\pi c_s^3}{6 G^{3/2}\rho_{\rm crit}^{1/2}},&
	M_{\rm crit,\Sigma}\sim \frac{c_s^4}{G^2 \Sigma_{\rm crit}},
	\end{eqnarray}
	where $\rho_{\rm crit}$ and $\Sigma_{\rm crit}$ are the critical densities for the isothermal-adiabatic transition.
	
	\item \textbf{Turbulent/sonic mass models:} A number of analytical theories derive the CMF and IMF from the properties of the turbulent medium, in which they form (e.g. \citealt{Padoan_Nordlund_2002_IMF,HC08,core_IMF,HC_2013}). In these models, both the CMF and IMF peaks are set by the ``sonic mass'' $M_{\rm sonic}$, namely the turbulent Jeans or Bonner-Ebert mass at the sonic scale ($R_{\rm sonic}$) below-which the turbulence becomes sub-sonic and therefore fails to generate large fluctuations (which seed fragmentation). The critical mass is
	\be
	M_{\rm crit,\,S} = M_{\rm sonic}\sim \frac{2 c_s^2 R_{\rm sonic}}{G},
	\ee
  where $R_{\rm sonic}$ is defined through the linewidth-size relation
  \be 
  \sigma^2_{\rm turb}(\lambda)=c_s^2 \frac{\lambda}{R_{\rm sonic}}.
  \ee
 In our calculations $\sigma_{\rm turb}^2$ is estimated from the simulations when a star particle forms by measuring the velocity dispersion (after subtracting the mean shear) between neighboring particles in a sphere of radius $\lambda$ (taken to be that which encloses the nearest $\sim32$ gas neighbours). 
	
	\item \textbf{Protostellar feedback models:} Although there are a number of ways newly-formed stars can regulate star formation, most studies have concluded that at the scale of the IMF peak (early protostellar collapse of $\sim 0.1\,\msun$ clouds) the most important self-regulation mechanism is radiative feedback from protostellar accretion \citep{Bate_2009_rad_importance, Krumholz_stellar_mass_origin, guszejnov_feedback_necessity}. This sets a unique mass and spatial scale within which the protostellar heating has raised the temperature to make the core Jeans-stable, terminating fragmentation. The resulting critical masses are 
	\begin{eqnarray}
	M_{\rm crit,B}\sim 0.5\left(\frac{\rho}{1.2\times 10^{-19}\,\mathrm{g/cm^3}}\right)^{-1/5}\left(\frac{L_{*}}{150\,\solarluminosity}\right)^{3/10}\,\solarmass,\\
	M_{\rm crit,K}\sim 0.15\left(\frac{P/k_B}{10^{6}\,\mathrm{K/cm^3}}\right)^{-1/18}\solarmass
	\end{eqnarray}
	where $L_{*}$ is the average luminosity of accreting protostars and $P$ is the pressure. These different formulas come from \citet{Bate_2009_rad_importance} and  \citet{Krumholz_stellar_mass_origin}, respectively; the differences are due to the detailed uncertainties treating radiation. However for our purposes they give {\em nearly identical} results, so we will focus on the model from \citet{Krumholz_stellar_mass_origin}.
\end{itemize}


\begin{table*}
	\centering
		\begin{tabular}{cccccccc}
		\hline\\
		\multirow{4}{*}{\bf Model} & \multirow{4}{*}{$M_{\rm crit}$} & \multirow{4}{*}{\bf Reference} & \multicolumn{5}{c}{\bf Galactic IMF variation ($\sigma_{M_{\rm crit}}$) [dex]} \\
        \cline{4-8}
        & & & {\bf m12i} & {\bf m12i} & {\bf m12i+MHD} & {\bf m12f} & {\bf m12m}\\
        & & & ({$56000\,\solarmass$}) & ({$7000\,\solarmass$}) & ({$7000\,\solarmass$}) & ({$7000\,\solarmass$}) & ({$7000\,\solarmass$}) \rule{0pt}{2.6ex}\\
        & & & ({$1000\,{\rm cm^{-3}}$}) & ({$1000\,{\rm cm^{-3}}$}) & ({$1000\,{\rm cm^{-3}}$}) & ({$1000\,{\rm cm^{-3}}$}) & ({$1000\,{\rm cm^{-3}}$}) \rule{0pt}{2.6ex}\\
		\hline
        \hline
		Jeans Mass & $\propto T^{3/2}\rho^{-1/2}$ & \citealt{BateBonell2005} & 0.60 & 0.54 & 0.61 & 0.56 & 0.65 \\
        \hline
		Turbulent/Sonic Mass & $\propto T R_{\rm sonic}$ & \citealt{core_IMF} & 0.91 & 0.96 & 0.86 & 0.91 & 0.84\\
        \hline
		Opacity-limited, $\rho_{\rm crit}$ & $\propto T^{3/2}$ & \citealt{Jappsen_EQS_ref}& 0.63 & 0.53 & 0.58 & 0.54 & 0.61\\
        \hline
		Opacity-limited, $\Sigma_{\rm crit}$ & $\propto T^{2}$ & \citealt{Bate_2009_rad_importance}& 0.81 & 0.70 & 0.75 & 0.73 & 0.81\\
        \hline
		Protostellar Heating & $\propto \left(\rho T\right)^{-1/18} $  & \citealt{Krumholz_stellar_mass_origin}& 0.030 & 0.026 & 0.031 & 0.027 & 0.031 \\
		\hline
		\end{tabular}
        \vspace{-0.1cm}
 \caption{Rows: Different IMF models compared in this paper, each with the predicted scaling of the IMF turnover mass $M_{\rm crit}$ with initial parent cloud properties (\S~\ref{sec:methods.imf.models}), reference, and the predicted $1\sigma$ dispersion in $\log_{10}(M_{\rm crit})$ across the galaxy at present-day (averaging Fig.~\ref{fig:physical_var} over all galacto-centric radii). We measure $\sigma_{M_{\rm crit}}$ from five simulations: galaxies {\bf m12i}, {\bf m12f}, and {\bf m12m} are three distinct Milky Way-mass ($\sim 10^{12}\,\msun$) halos which produce similar disky, Milky Way-like galaxies (stellar mass $\sim 0.5-1\times10^{11}\,\msun$), but have different formation histories (see Fig. \ref{fig:all_galaxy_views} for visualizations and \citealt{Hopkins2017_FIRE2} for details). For each we label the mass resolution (in $\msun$) and minimum density $n_{\rm crit}$ for creation of stellar sink particles. For {\bf m12i}, we compare two alternative runs: one at lower resolution, and one including magnetic fields ({\bf m12i+MHD}). The predicted IMF variation is remarkably robust across all these simulations.
 \label{tab:SF_models}} \vspace{-0.5cm}
\end{table*}

\vspace{-0.5cm}
\section{Results and Discussion}\label{sec:results}

\begin{figure}
\begin {center}
\includegraphics[width=\linewidth]{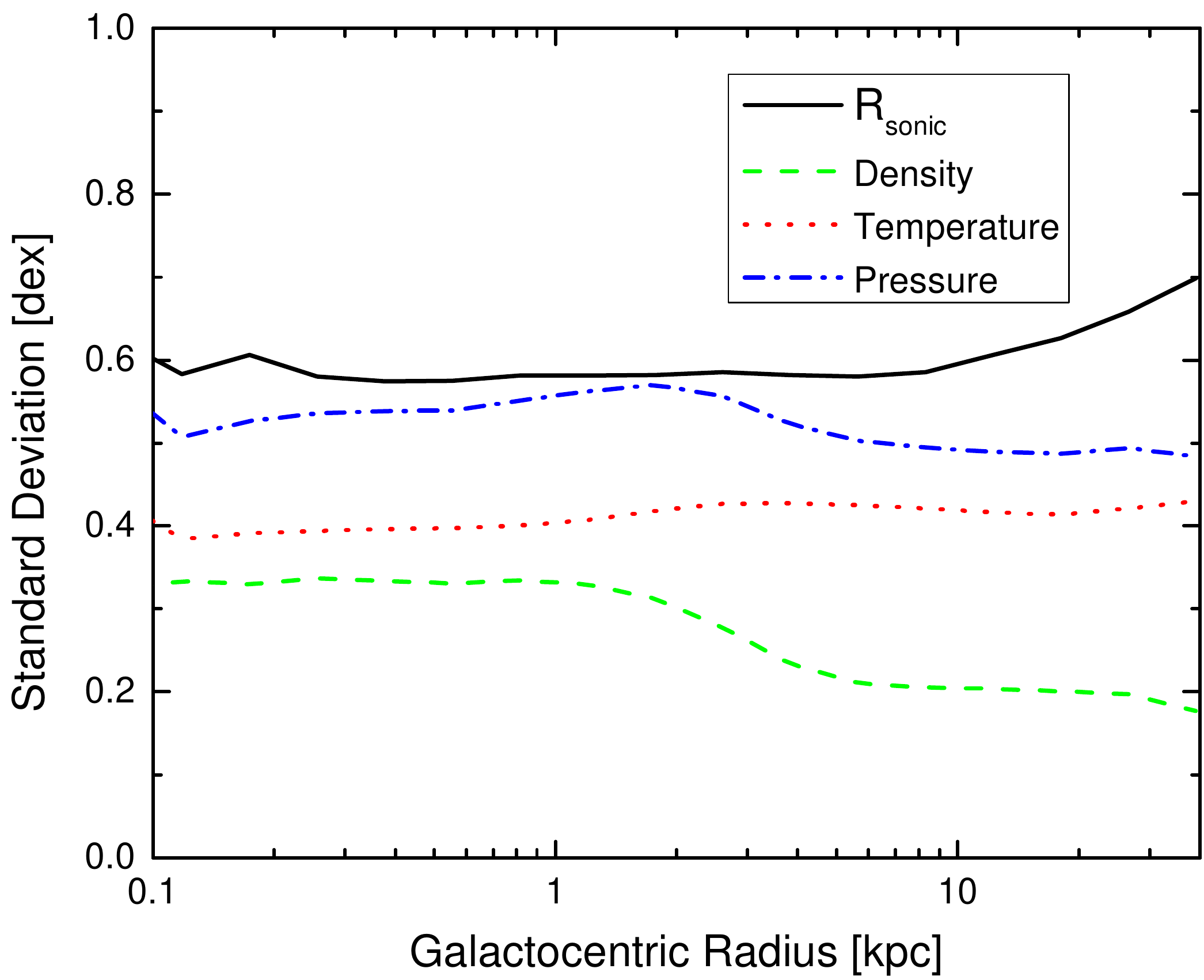}
\vspace{-0.2cm}
\caption{Standard deviation in star-forming progenitor cloud properties (measured {\em at the time of star formation}, as in Fig.~\ref{fig:n_T_hist}), across the progenitor clouds of all stellar sink particles which reside at a given present-day galacto-centric radius (in our fiducial \emph{m12i} run with $56000\,\solarmass$ resolution; however the dependence on radius is weak and all our simulations in Table~\ref{tab:SF_models} give similar results). Note that this is \emph{not} the variation of present-day star forming clouds at different radii, as stars at some present-day radius could have formed at wildly different times and positions (for example, at high redshift in a more gas-rich disk with much larger pressures and densities). Thermodynamic and turbulent progenitor-cloud properties vary by $\sim 0.3-0.5$\,dex; this implies large IMF variations for any model which has a strong dependence on these quantities.
\label{fig:physical_var}}
\vspace{-0.5cm}
\end {center}
\end{figure}

\begin{figure}
\begin {center}
\includegraphics[width=0.9\linewidth]{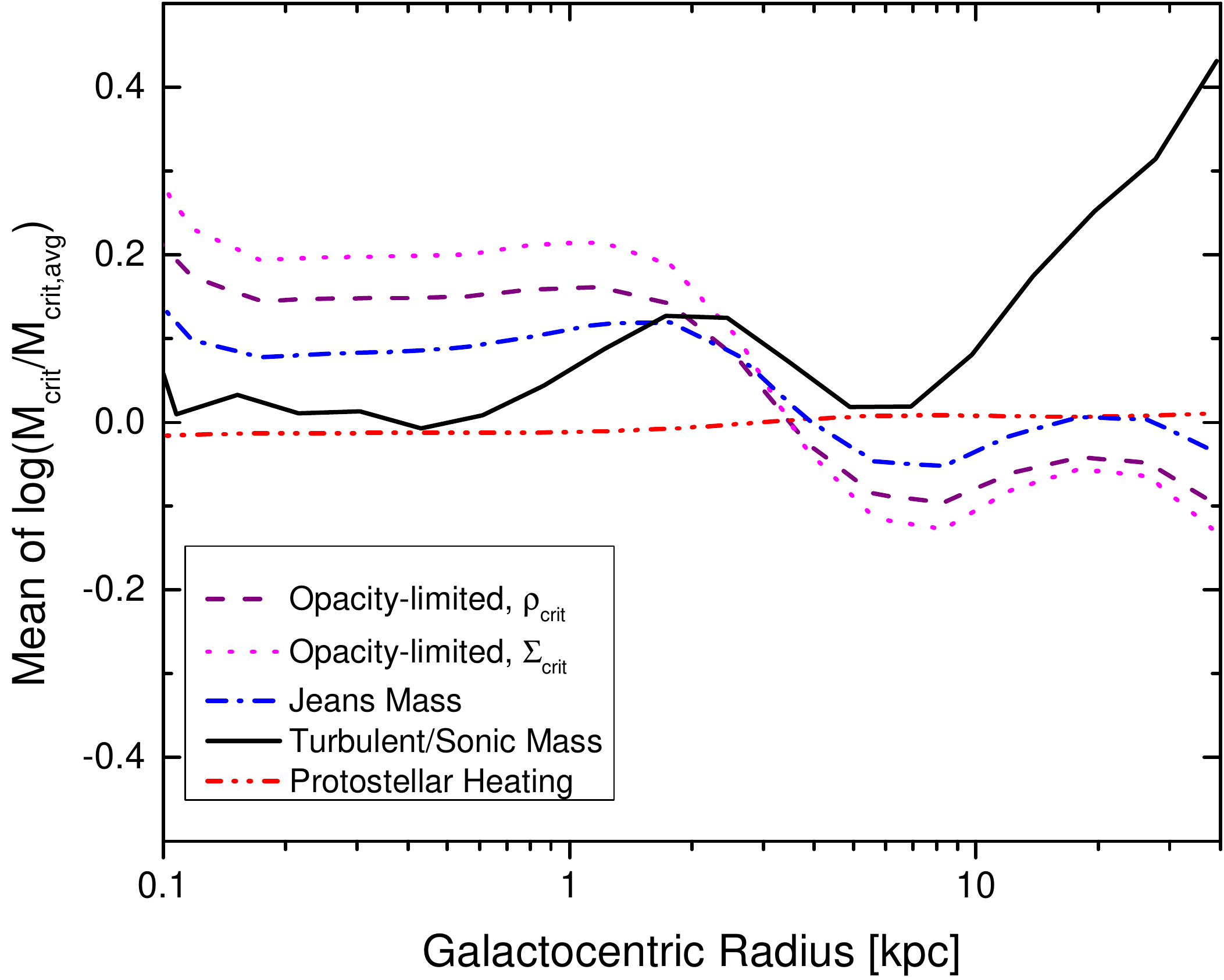}\\
\includegraphics[width=0.9\linewidth]{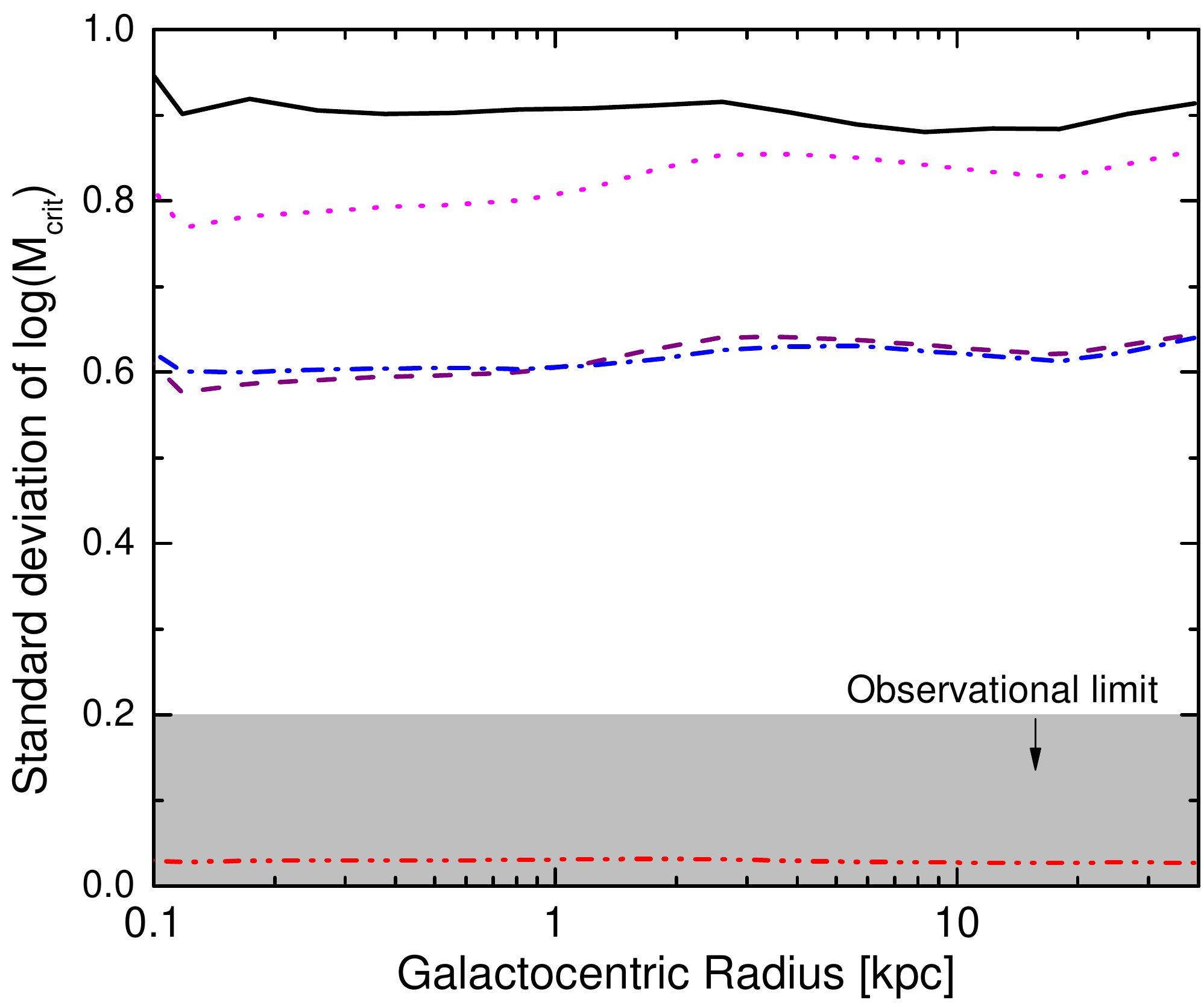}
\caption{Mean (top) and standard deviation (bottom) of the IMF turnover mass $M_{\rm crit}$ normalized its galactic average ($M_{\rm crit, avg}$) at different galacto-centric radii (in our fiducial \emph{m12i} run with $56000\,\solarmass$ resolution). We compare the IMF models in Table \ref{tab:SF_models} and the observationally allowed range of scatter in the IMF across the Milky Way, from \citet{IMF_review}. In these simulations, only models accounting for protostellar heating avoid strongly over-predicting the scatter in MW IMFs. The models are shown here for the same example galaxy in Fig.~\ref{fig:physical_var}, but we obtain very similar results for each of the five simulated galaxies in Table~\ref{tab:SF_models}.
\label{fig:turnover_var}}
\vspace{-0.5cm}
\end {center}
\end{figure}

Fig.~\ref{fig:physical_var} shows that there is significant variation the properties of the progenitor GMC complexes which formed stars that ultimately end up at a specific galacto-centric radius. We stress that this is not the variation of properties in {\em present-day} star-forming clouds, but includes all variations in time as well: if the galaxy progenitor was gas-rich (gas fraction $\sim 1/2$) at $z\sim1-2$ for example, then the midplane gravitational pressure ($\sim G\,\Sigma_{\rm gas}^{2}$) would have been a factor $\sim 100$ larger than in the galaxy today. Fig.~\ref{fig:turnover_var} shows that this, in turn, produces large IMF variations in all models here except those accounting for protostellar heating. Such variations ($>0.5\,$dex in $M_{\rm turnover}$) are strongly ruled-out by observations \citep{IMF_review}. Note that these results are robust to variations in simulation parameters (see Table \ref{tab:SF_models}).

The variations in the IMF predicted by some of the simple models here (e.g.\ the Jeans-mass models) have often been substantially underestimated in previous work in the literature. In analytic models of the IMF (see references in Sec. \ref{sec:intro}) or galaxy-scale models which fail to resolve individual ``parent clouds'', but post-process the entire galaxy (with $>$\,kpc-scale resolution) to determine an IMF \citep{Narayanan_2012_jeans_extragalactic,Hopkins_CMF_var,Blancato_2016_Illustrus_IMF_var}, it is commonly assumed that all star-forming clouds are uniformly at the same isothermal temperature (e.g.\ $T=10\,$K at {\em all} densities), virial parameter, and lie exactly on the same linewidth-size relation. For example, if the gas {\em at all densities and all cosmic times} had exactly the same temperature, then the variation in the IMF for the opacity-limited EOS models would vanish (all clouds and cores lie on exactly one adiabat). This assumption is not correct, however, as even in the present-day MW (e.g.\ fixed redshift and galaxy properties) both GMC and clump temperatures \citep[e.g.][]{Bergin_Tafalla_2007,MillsHotAmmoniaGalacticCenter,SanchezCoreTemperatureRangesHundredsKelvin,OttGalacticCenterAmmoniaTempVariations,Nishimura_GMC_temp_2015}\footnote{Note that \protect\cite{Nishimura_GMC_temp_2015} only focuses on Orion A and B so these results are not necessarily representative of the entire MW. Also, as our stars form primarily around $n_{\rm crit}=1000\,{\rm cm}^{-3}$, the average temperature of the progenitor clouds is higher than the observed GMCs because they have not reached the cooler, higher-density fully-molecular phases (see Figs.~\ref{fig:n_T_hist_z0}-\ref{fig:n_T_hist}).} and virial parameters \citep[e.g.][]{Kauffmann_Pillai_2013,Svoboda_2016} vary substantially. As expected variations are more pronounced in other nearby dwarf or star-forming galaxies \citep{GorskiAmmoniaTempsNearbySFGals50to130K,TangLMCstarformingGasAtRho1000hasTof16to63K} or redshift $z\gtrsim 1-2$ galaxies and starburst systems \citep[see e.g.][]{Ott2011Arp220GasIs200Kelvin,Gonzalez2dexVariationInGMCTempsInStarbursts,NarayananSLEDmodelsNearbyGal,MagnumLargeVariationsAmmoniaAndDustTempsofNormalSFGals,Miyamoto300KammoniainSeyfert,ZschaechnerArp220GasTemps100to300K}, which are better analogues to the progenitors where many of the stars in the present-day Galaxy formed. 

It is certainly possible that we (and these observations) have over-estimated the range of temperatures of GMCs in different environments. But the strong temperature sensitivity of the EOS models (e.g.\ $\propto T^{2}$) means that the temperature of {\em all} progenitor clouds, at all redshifts and in all progenitor galaxies, which formed stars in today's MW, would have to lie within a scatter of just $\sim 20\%$ in temperature (smaller than that observed in just solar-neighborhood clouds) in order to avoid exceeding the allowed IMF variation in the MW.

Moreover the linewidth-size relation is observed to vary systematically, both within the MW and galaxy-to-galaxy, with high-redshift galaxies (the progenitors of the MW) differing by more than an order of magnitude \citep[see e.g.][and references therein]{Swinbank2011linewidthsize,Swinbank2015linewidthsize,Canameras2017highZlinewidthsize}. Even if temperature variations are neglected entirely, in the ``Turbulent/Sonic Mass'' models the turnover mass is proportional to the {\em square} of the deviation ($(\sigma_{\rm cloud}/\langle\sigma[R]\rangle)^{2}$) of each cloud from the linewidth-size relation \citep{core_IMF, HC08}, but these deviations are observed to be $\sim 0.3-0.5\,$dex even within the MW at present-time \citep{Bolatto_2008} implying $> 0.6$\,dex scatter. Likewise the density dependence in ``Jeans Mass'' models predicts $>0.3$\,dex scatter even if all temperature variations, time variations, and progenitor-galaxy variations are neglected (e.g. if we use only the scatter in cloud densities observed in the solar neighborhood of the MW at the present instant). 

Recall, the cloud properties we use to predict the IMF are measured at a density scale of $\sim 1000\,{\rm cm^{-3}}$ and mass scale $\sim 7000-56000\,\msun$. Obviously the clouds must continue to evolve and fragment to form actual stars -- this is what our cloud-scale IMF models attempt to model. One might wonder, however, whether during this process some of the scatter might be reduced (if, for example, the clouds all converged to the same temperature eventually, owing to some additional physics). In the opacity-limited models, the equation of state (EOS) is specified (generally the cloud cools with $T\propto \rho^{-0.3}$ to some density, becomes approximately isothermal, then becomes adiabatic above the opacity-limit density), so this is already built into the model explicitly. In the ``Jeans Mass'' or ``Turbulent/Sonic Mass'' models, we have implicitly assumed an isothermal EOS within each cloud so their temperature was assumed to be constant throughout their evolution and set by the initial conditions. One might, therefore, consider a more complicated version of these models (different from the simple scalings used thus far). Let us assume star formation occurs above some critical density $\rho_{\rm crit}$ and the gas follows a polytropic EOS with index $\gamma$. The critical mass (Table~\ref{tab:SF_models}) will then depend on $T_{\rm crit}(\rho=\rho_{\rm crit})=T_{\rm cloud}\,(\rho_{\rm crit}/\rho_{\rm cloud})^{\gamma-1}$, as $M_{\rm crit} \propto T_{\rm crit}^{\alpha}$ where $\alpha=3/2,\,1$ for the Jeans and Turbulent/Sonic models, respectively. Some simple algebra then gives us logarithmic variance in $M_{\rm crit}$, $S_{\log{M_{\rm crit}}} = \alpha\,(S_{\log{T_{\rm cloud}}} + \,[\gamma-1]\,S_{\log{\rho_{\rm cloud}}} )$. Putting in the actual values (Fig.~\ref{fig:physical_var}) this gives a dispersion $\sigma_{\log{M_{\rm crit}}} \approx 0.6,\,0.4$\,dex for the Jeans and Turbulent/Sonic models (for any $\gamma\sim 0.5-1.5$). This reduces the predicted IMF variation, but still leaves it far larger than observed.

Thus we have shown that {\em some} additional physics on cloud or sub-cloud scales must be accounted for to reconcile the predictions of the ``no-feedback'' IMF models with the (weak) IMF variations observed in the MW. The ``protostellar heating'' models represent one physically-motivated class of models that do exactly this. Of course there may be others, but, broadly-speaking, they would need to either (a) strongly reduce the level of dependence of the predicted IMF on cloud properties (as the protostellar heating models do), or (b) strongly reduce the variation in GMC-scale properties predicted across cosmic time in the progenitor galaxies that form the MW. The latter is not impossible but seems to contradict the direct observations cited above, showing large variations in cloud properties in distant galaxies.

\subsection{Caveats}

Of course detailed, complex simulations (like the cosmological FIRE runs we are using) employ a large number of approximations to make problems numerically tractable. Although these simulations have been extensively vetted numerically \citep[for details see][]{Hopkins2017_FIRE2} some caveats worth noting include:
\begin{itemize}
	\item Our analysis uses a somewhat arbitrary $n_{\rm crit}=1000\,\rm cm^{-3}$ minimum density threshold for star formation, based on numerical considerations. Using a much higher threshold would require much greater mass resolution (or else it would introduce severe numerical artifacts), which is not computationally feasible (these are the highest-resolution cosmological simulations of MW-mass galaxies ever run, at present). However, within the range we can probe, our results do not appear to depend sensitively on the density threshold or other numerical criteria for star formation.\footnote{In \citet{Orr_FIRE_KS}, we show the results of a number of simulations where we re-run our {\bf m12i} galaxy from $z\approx 0.1-0.0$, as in Fig.~\ref{fig:mass_functions}, but vary the numerical SF criteria. This includes changing the minimum SF density (from $\sim 10-1000\,{\rm cm^{-3}}$), removing requirements that the gas be molecular and/or self-gravitating, and changing the efficiency per free-fall time with which gas that meets this criteria will turn into stars (from $\sim1-100\%$). We have re-run our analysis, restricted to just those stars formed over the period the simulations were re-run, and find these changes do not significantly influence the predicted IMF variations.}
    
	\item In the simulations, gas elements are replaced by star particles instantaneously once all star formation criteria and timescales are satisfied, so star formation happens in discrete steps. In the large GMCs where most stars form ($\sim 10^{6}-10^{7}\,\msun$), this means that the first generation of stars formed can continue to alter the GMC properties while subsequent star formation occurs. However star formation in the smallest GMCs will be artificially ``abrupt'' (although GMCs with masses this low contribute negligibly to the variation in the IMF). 
    
    \item Feedback processes from low-mass stars, e.g.\ proto-stellar outflows, are not explicitly included in the simulations. We only consider the effects of massive stars, which dominate on GMC scales provided there are sufficient stars to sample the IMF.
    
	\item The turbulent velocity dispersion in the code is calculated from using a kernel interpolation over the the relative velocities between the nearest $\sim 32$ resolution elements, after subtracting the coherent shear and contraction/expansion terms. This means that for very small GMCs with masses $\lesssim 10$ times the resolution, internal motions are not well-resolved. In this limit the general tendency is to {\em under}-estimate the turbulent velocity dispersions (see e.g.\ the detailed turbulence studies in \citealt{Hopkins2015_GIZMO}). But again, these do not contribute significantly in our predictions.
    
	\item The simulations do not explicitly follow non-equilibrium chemistry (e.g.\ molecular hydrogen formation/destruction), instead relying on pre-tabulated equilibrium cooling rates as a function of density, temperature, metallicity, and the strength of the local radiation field in several bands. It was shown by \cite{Hopkins_2012_galaxy_structure} that these approximations have little to no effect on galactic star formation properties but they could conceivably alter the scatter in small-scale cloud properties. 
\end{itemize}

\section{Conclusions}

In this paper we explore the application of broad classes of IMF models to high-resolution fully-cosmological galaxy formation simulations. Stars at a some present-day location might have formed at very different times and places, in an environment radically different from today's MW: only by using a cosmological simulation instead of local simulations or observations can we predict the properties of their progenitor star-forming clouds at these times and places, and therefore use these models to predict e.g.\ the variations in the IMFs predicted for old stellar populations in the present-day galaxy. This also provides an important consistency and validity check for future attempts to incorporate these IMF models into such simulations {\em dynamically}, as stellar feedback plays a critical role in the simulations and it, obviously, depends on the IMF.

In summary, we find that only models accounting for protostellar heating produce sufficiently weak IMF variations, in these simulations, to be compatible with observations. This discrepancy is not obvious in many previous studies (either analytic or idealized single-cloud simulations) as they artificially assume all clouds (at all locations and cosmic times) have the same temperature and obey the same linewidth-size relation (without scatter or systematic variation), whereas observations find significant variations in molecular gas temperatures and velocity dispersions (both within the MW and in nearby and high-redshift star-forming galaxies, which may more closely resemble the MW progenitors where these stars formed). 

The protostellar heating models, on the other hand, actually predict IMF variations significantly below the observational upper limits (see Fig.~\ref{fig:turnover_var}). Additional sources of variance are therefore easily accommodated in these models, such as those that should come from a combination of (a) stochastic statistical sampling effects (see \citealt{IMF_review}; these may be especially important in small clouds such as Taurus which are not resolved by our simulations, see \citealt{Kraus_2017_Taurus}), (b) measurement uncertainties, or (c) additional physics not accounted for by the model (e.g.\ bursty accretion or other physics may modify the radiative efficiency and heating effects of protostars, introducing some IMF variation). 

In future work, we will examine whether the protostellar heating models considered in this study should produce observably-large IMF variation under more extreme conditions. Preliminary comparison of single-cloud conditions in \citet{guszejnov_feedback_necessity} suggests these models can produce as much as factor $\sim 2$ shifts in the turnover mass under extreme starburst conditions analogous to Arp220, but this needs to be explored in more detail. We will also explore in more detail IMF shape variations, the predicted IMF in different sub-regions of the galaxy (e.g.\ the galactic nucleus), and the IMF in specific populations (e.g. metal-poor globular clusters versus present-day stellar populations).

\vspace{-0.5cm}
\acknowledgments
Support for PFH, DG, and XM was provided by an Alfred P. Sloan Research Fellowship, NASA ATP Grant NNX14AH35G, and NSF Collaborative Research Grant \#1411920 and CAREER grant \#1455342. Numerical calculations were run on the Caltech compute cluster ``Zwicky'' (NSF MRI award \#PHY-0960291) and allocation TG-AST130039 granted by the Extreme Science and Engineering Discovery Environment (XSEDE) supported by the NSF. 

\bibliographystyle{mnras}
\bibliography{bibliography}

\end{document}